\documentclass[sigconf]{acmart}

\usepackage{tikz}

\usepackage{pifont}
\PassOptionsToPackage{table,dvipsnames,svgnames,x11names}{xcolor}
\usepackage[svgnames,table]{xcolor}
\usepackage{colortbl}

\definecolor{Green}{rgb}{0.0,0.5,0.0}        %
\definecolor{DarkGoldenrod}{RGB}{184,134,11} %

\usepackage{array}
\newcolumntype{C}[1]{>{\centering\let\newline\\\arraybackslash\hspace{0pt}}p{#1}}

\usepackage{listings}
\usepackage{enumitem}

\newcommand{\thiswork}{OmniSim}
\newcommand{\numbenchmarks}{eleven}

\newcommand*\circled[1]{\raisebox{.4pt}
                    {\tikz[baseline=(char.base)]{
            \node[shape=circle,draw,inner sep=1pt, style={fill=black, text=white}, scale=0.75] (char) {\textbf{#1}};}}}
\newcommand*\emptycirc[1][1ex]{\tikz\draw (0,0) circle (#1);} 

\newcommand*\fullcirc[1][1ex]{\tikz\fill (0,0) circle (#1);}

\usepackage{pifont}
\usepackage{multirow}

\newcommand{\para}[1]{\vspace{2pt}\noindent{\bfseries{$\blacktriangleright$ #1}}}

\copyrightyear{2025} 
\acmYear{2025} 
\setcopyright{cc}
\setcctype{by}
\acmConference[MICRO '25]{58th IEEE/ACM International Symposium on Microarchitecture}{October 18--22, 2025}{Seoul, Republic of Korea}
\acmBooktitle{58th IEEE/ACM International Symposium on Microarchitecture (MICRO '25), October 18--22, 2025, Seoul, Republic of Korea}\acmDOI{10.1145/3725843.3756033}
\acmISBN{979-8-4007-1573-0/2025/10}

\begin{document}

\title{OmniSim: Simulating Hardware with C Speed and RTL Accuracy for High-Level Synthesis Designs}

\author{\normalsize Rishov Sarkar}
\affiliation{
  \institution{\normalsize Georgia Institute of Technology}
  \city{\normalsize Atlanta}
  \country{\normalsize GA}
}
\email{rishov.sarkar@gatech.edu}

\author{\normalsize Cong Hao}
\affiliation{
  \institution{\normalsize Georgia Institute of Technology}
  \city{\normalsize Atlanta}
  \country{\normalsize GA}
}
\email{callie.hao@gatech.edu}

\begin{abstract}
High-Level Synthesis (HLS) is increasingly popular for hardware design using C/C++ instead of Register-Transfer Level (RTL). To express concurrent hardware behavior in a sequential language like C/C++, HLS tools introduce constructs such as infinite loops and dataflow modules connected by FIFOs (first-in first-out). While these constructs can \textit{represent} concurrency, efficiently and accurately \textit{simulating} them at C level remains challenging.
First, without hardware timing information, functional verification typically requires slow RTL synthesis and simulation, as the current approaches in commercial HLS tools. Second, cycle-accurate performance metrics, such as end-to-end latency or throughput, also rely on RTL simulation.
No existing HLS tool fully overcomes the first limitation. For the second, prior work such as LightningSim partially improves simulation speed but lacks support for advanced dataflow features like cyclic dependencies and non-blocking FIFO accesses.

To overcome both limitations, we propose \textbf{\texttt{\thiswork{}}}, a framework that significantly extends the simulation capabilities of both academic and commercial HLS tools. First, \thiswork{} enables fast and accurate simulation of complex dataflow designs, especially those explicitly declared unsupported by commercial tools. It does so through sophisticated software multi-threading, where threads are orchestrated by querying and updating a set of FIFO tables that explicitly record exact hardware timing of each FIFO access.
Second, \thiswork{} achieves near-C simulation speed with near-RTL accuracy for both functionality and performance, via flexibly coupled and overlapped functionality and performance simulations.

We demonstrate that \thiswork{} successfully simulates \textbf{\numbenchmarks} designs previously unsupported by any HLS tool, achieving up to \textbf{35.9$\times$} speedup over traditional C/RTL co-simulation, and up to \textbf{6.61$\times$} speedup over the state-of-the-art yet less capable simulator, LightningSim, on its own benchmark suite.
\end{abstract}

\keywords{Dataflow Designs, Design Simulation, High-Level Synthesis}

\maketitle

\section{Introduction}
\label{sec:intro}

\begin{figure*}
    \centering
    \includegraphics[width=\linewidth]{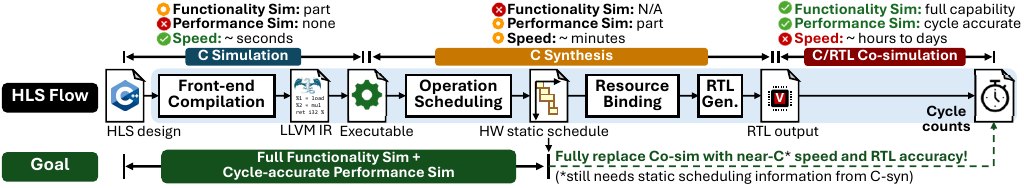}
    
    \vspace{-8pt}
    \caption{Overview of a representative HLS flow and the goal of this work.}
    \vspace{-4pt}
    \label{fig:hls}
\end{figure*}

\begin{figure}[t]
    \centering
    \includegraphics[width=\linewidth]{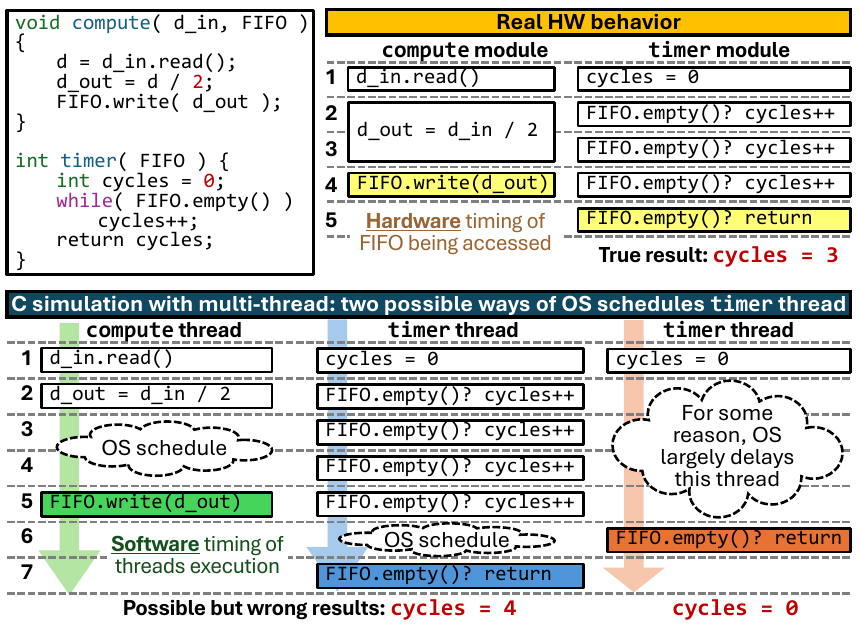}
    \vspace{-16pt}
    \caption{Accurately simulating hardware functionality and performance at C level is highly challenging due to the mismatch between sequential C and concurrent hardware: \textit{key actions (e.g., FIFO access) must be captured by their \textbf{hardware} timing, not \textbf{software} timing from thread scheduling.}}
    \label{fig:motivation}
\end{figure}

\para{Hardware design using HLS.}
High-Level Synthesis (HLS) has gained popularity in hardware design due to its significantly improved productivity, allowing users to develop hardware using high-level languages such as C/C++, which will be automatically synthesized into Register Transfer Level (RTL) code. When designing hardware, users typically care about two key aspects: (1) whether the design behaves functionally as expected, known as \textbf{functionality simulation}, and (2) how well the synthesized hardware performs in terms of latency or throughput, known as \textbf{performance simulation}.\footnote{Also referred to as functionality verification, timing simulation, or performance testing by different HLS tool vendors.}

As illustrated in Fig.~\ref{fig:hls}, a typical HLS flow consists of three stages: C simulation, C synthesis, and C/RTL co-simulation. C/RTL co-simulation offers the highest fidelity with cycle-level validation, but is notoriously slow, often requiring hours or days, undermining the productivity benefits of HLS. At the other extreme, C simulation is fast and lightweight, but provides only limited functional validation and no performance insights. C synthesis lies in between: after generating RTL, the tool produces static performance estimates, which unfortunately are often inaccurate or even unavailable, especially with variable loop bounds or complex control flow.

While much recent work, such as Verilator~\cite{Verilator} and FireSim~\cite{karandikar2018firesim}, focuses on speeding up \textit{RTL} simulation, our focus is on an earlier point in the design cycle: \textit{simulating HLS designs before RTL is even generated, for both functionality and performance}. Our goal is to deliver RTL-level accuracy at near-C simulation speed, enabling rapid design-space exploration without waiting for RTL generation.

\para{Challenges of C-level simulation.}
Current HLS tools can simulate functionality at the C level for \textit{some} designs, but their performance estimation is often inaccurate due to the lack of run-time details such as actual loop counts, branch behavior, and FIFO stalls~\cite{lightningsim,lightningsimv2,flash,fastsim}.
The challenges become greater for \textit{more complex designs}. To express concurrent hardware behavior in sequential C/C++, HLS tools rely on special constructs such as infinite loops and dataflow (streaming) modules connected by non-blocking FIFOs. These constructs can \textit{represent} a wide range of hardware designs, but they make it particularly difficult to correctly \textit{simulate} them at the C level, both for functionality and for performance.

\circled{1} \textbf{Functionality simulation.}
Certain common RTL scenarios are poorly supported by C-level simulation. For example, a timer module that tracks the cycle count of another module, or a network router that dynamically changes output ports depending on congestion, must be implemented as \textbf{dataflow designs} with \textit{non-blocking reads and writes} (see Sec.~\ref{sec:background}).
However, even commercial tools such as Vitis HLS~\cite{vitis-hls} and Catapult HLS~\cite{catapult-hls} struggle to simulate such behavior correctly. Their manuals explicitly warn that C-level simulation may produce incorrect behavior, forcing users to rely on RTL simulation, which is both slow and tedious.\footnote{\textbf{Catapult manual:} \textit{``If the top level of your design contains non-blocking IO, the tool may not produce results as expected.''} \textit{``While the synthesis of a non-blocking read is trivial, the resulting hardware will rarely, if ever, match the functionality of the original C code; all testing done at the C level would need to be re-done in the RTL.''}}%
\textsuperscript{,\,}%
\footnote{\textbf{Vitis HLS manual:} \textit{``Using non-blocking APIs can lead to non-deterministic behavior which cannot be fully validated during C-simulation and requires an RTL test bench to test it exhaustively.''}}

\circled{2} \textbf{Performance simulation}.
Beyond functionality simulation, accurately predicting performance, such as cycle counts or throughput, at the C level is even more challenging. This is because performance depends not only on correct functional behavior (already difficult to achieve at this level) \textit{but also} on hardware-specific details such as scheduling, pipelining, and finite-state machine (FSM) transitions. Similarly, commercial HLS tools lack satisfactory solutions, either providing rough and inaccurate performance estimates after full synthesis, or fall back to slow RTL simulation.\footnote{\textbf{Catapult manual:} \textit{``Non-blocking IO cannot be used for throughput-accurate simulation.''} \textit{``Latency testing using the source code is not commonly done (and is not even possible in C++-based designs), so there is no best practice for this yet.''}}%
\textsuperscript{,\,}%
\footnote{\textbf{Vitis HLS manual:} \textit{``During C simulation, streams have an infinite size. It is therefore not possible to validate with C simulation, and these methods can be verified only during RTL simulation when the FIFO sizes are defined.''}}

\circled{3} \textbf{Tightly coupled functionality and performance.}
The two challenges above both arise from the semantic gap between untimed, sequential C/C++ and the timed, concurrent nature of hardware. This gap creates a fundamental mismatch between \textit{hardware-level timing of actions} and the \textit{OS-controlled scheduling of software threads} during simulation.

To model concurrent hardware at the C level, particularly designs with multiple active modules with cyclic dependencies and/or infinite loops, it is natural to use C/C++ multi-threading to mimic simultaneous activity. However, in naive multi-threaded simulation, execution order is dictated entirely by the OS scheduler, which can distort timing and compromise correctness.

Fig.~\ref{fig:motivation} illustrates this with a simple example: a \texttt{timer} module monitors a \texttt{compute} module by polling a FIFO for results. The expected \texttt{cycles} value is three based on \textit{hardware timing}. However, if simulated naively with software threads, the timing of FIFO accesses is governed by the OS’s scheduling—i.e., \textit{software timing}—leading to incorrect results, as shown in the two cases in Fig.\ref{fig:motivation}.

This example highlights two key implications. First, correct simulation must preserve \textit{true hardware timing}, independent of OS scheduling, by explicitly orchestrating thread execution and tracking cycles (e.g., ensuring accurate FIFO full/empty status at every cycle). Second, in many dataflow designs, correct functionality depends directly on correct timing, making functionality and performance simulation inherently inseparable and tightly coupled.

\para{Existing work.}
For complex dataflow designs (e.g., with non-blocking FIFO access) like the one presented in our motivating example, existing HLS tools—including commercial ones—fail to correctly simulate functionality, let alone performance. For non-dataflow designs or simpler ones, while functionality simulation is often straightforward, achieving fast yet cycle-accurate performance simulation remains a significant challenge. Various tools such as FLASH~\cite{flash}, FastSim~\cite{fastsim}, and LightningSim~\cite{lightningsim,lightningsimv2} have been developed as faster alternatives to RTL simulation, but they support only a limited subset of HLS designs.

\para{Our approach.} 
To address the limitations of existing HLS simulation, we propose \thiswork{} to: (1) \textbf{extend C-level simulation capability of HLS tools} by enabling both functionality and performance simulations for those \textit{complex dataflow designs} that are currently unsupported or considered infeasible; (2) \textbf{achieve near-C speed while maintaining RTL accuracy}, for both simple and complex designs. We propose the below key technologies:
\begin{itemize}[leftmargin=*]
\item \textbf{Sophisticated orchestration for software multi-threading.} 
First, to simulate concurrent behaviors of multiple active hardware modules, especially those with cyclic dependencies, we propose software-based multi-threading, where the functionality of each active module is simulated by one software thread.
Second, to align software thread execution with actual hardware timing, we introduce a dedicated performance simulation thread with an orchestration mechanism among the threads, including: thread requests, queries, and FIFO tables, to accurately track hardware-level timing despite arbitrary OS scheduling. This ensures functional correctness that precisely matches RTL behavior—something commercial tools explicitly declare infeasible.

\item \textbf{Flexibly coupled and overlapped functionality and performance simulation.} Through multi-thread orchestration, \thiswork{} flexibly separates and re-integrates functionality and performance simulation on demand. This is not only essential to expanding HLS simulation capability, but the overlapped execution also delivers significant speedup compared to the state-of-the-art (but less capable) simulator, LightningSim.

\end{itemize}

\noindent Other contributions are highlighted as follows:

\begin{itemize}[leftmargin=*]

\item \textbf{Taxonomy and classification of dataflow designs.} We present the first systematic, in-depth analysis of dataflow designs to determine whether they can be naively simulated using standard C/C++. We categorize them into \textbf{Type A, B,} and \textbf{C}, and explain the fundamental simulation challenges.

\item \textbf{Deadlock detection and incremental simulation.}  
\thiswork{} includes a crucial feature to detect deadlocks immediately as they occur, which are common in dataflow designs, without causing the simulation to hang. It is also able to maximally reuse previous simulation progress when FIFO sizes are modified, avoiding a complete restart and saving simulation time.

\item \textbf{Evaluation.} We design a benchmark suite consisting of \numbenchmarks{} C programs that cannot be correctly simulated at C-level by previous HLS tools. All of them can now be successfully simulated by \thiswork{} while being 30.7\texttimes{} faster than C/RTL co-simulation. We also demonstrate superior performance over the state-of-the-art HLS simulator, achieving up to 6.61\texttimes{} speedup. Furthermore, the incremental simulation capability of \thiswork{} shows \textit{four orders of magnitude} speedup over full simulation.

\end{itemize}

\vspace{-4pt}
\section{Background}
\label{sec:background}

\subsection{High-Level Synthesis Simulation}
\label{sec:how-hls-works}

As depicted in Fig.~\ref{fig:hls}, most HLS tools (using Vitis HLS as an example) have three stages, where the output of each stage presents clear tradeoffs in terms of functionality and performance simulation:

\begin{itemize}[leftmargin=*]
    \item \textbf{C Simulation (C-sim)}. HLS front-end compiles the source code into intermediate representation (IR), such as LLVM IR, and executes a standard C/C++ binary to verify functional correctness. \textit{C-sim executes in seconds and enables partial functional simulation, but it lacks hardware semantics and provides no performance estimates.} In addition, as already shown in Fig.~\ref{fig:motivation} (Sec.~\ref{sec:intro}), existing C-sim cannot correctly simulate complex dataflow designs.

    \item \textbf{C Synthesis (C-syn)}. After scheduling and binding, finite state machines (FSMs) are generated to represent control flow, and operations are scheduled and bound to specific hardware resources. The output is a static hardware schedule that captures module dependencies, concurrency, and pipeline parallelism. \textit{C-syn provides a rough estimate of static latency but does not model dynamic behaviors (e.g., non-blocking) and does not validate functionality.}

    \item \textbf{C/RTL Co-simulation (Co-sim)}. After RTL generation, testbenches are also generated to enable cycle-accurate simulation. Co-sim\footnote{Terminology in Vitis HLS; Catapult uses SCVerify with similar capabilities.} verifies both functional correctness and performance, reporting detailed hardware performance in terms of cycle counts. \textit{However, it is significantly slower—often requiring hours or even days to simulate a single testbench.}
\end{itemize}

\vspace{-4pt}
\subsection{Dataflow Designs}

Dataflow designs are prevalent in many domain-specific accelerators~\cite{tapa,abi2023inr,lerner2024data,basalama2025stream,nowatzki2017stream,kim2016dataflow,sarkar2023flowgnn}, enabling tasks to execute concurrently and communicate via FIFO channels for high performance. Beyond accelerators, they are also effective in representing real-world hardware designs such as multi-core, out-of-order, and network switching/routing. In HLS, users can define parallel tasks as functions, which are synthesized into hardware modules connected by FIFOs.

\subsubsection{FIFO Access: Blocking and Non-Blocking}

FIFO streams have two access patterns: \textit{blocking} (B) and \textit{non-blocking} (NB). Blocking accesses either complete immediately if the operation is feasible or stall the task until it becomes so. For example, a task writing to a full FIFO will stall until space becomes available (i.e., an element is consumed), and similar for a read operation from an empty FIFO.

In contrast, non-blocking accesses never stall the hardware, but simply return a Boolean value indicating the success or failure of the read/write operation. For instance, a non-blocking read (\texttt{read\_nb()} in Vitis HLS) returns \texttt{true} and pops an element if the FIFO is not empty; otherwise, it returns \texttt{false} immediately without waiting. Methods such as \texttt{empty()} and \texttt{full()} are also commonly used to inspect FIFO status, typically in conjunction with \texttt{if-else} branches.

Non-blocking access is critical in many real-world applications. For example, a real-time video processor must handle frames as they arrive; a non-blocking pipeline allows frames to be dropped under heavy load, avoiding backpressure and enabling timely processing of subsequent frames. Similarly, in network switches, non-blocking packet handling enables dynamic rerouting when certain processing paths are congested. More broadly, non-blocking behavior is fundamental to architectural features such as out-of-order CPU execution and non-blocking cache miss handling.
However, despite its importance, simulating non-blocking behavior at the C level is challenging and generally unsupported by commercial HLS tools.

\subsubsection{Dataflow Graphs: Acyclic and Cyclic}

When connected via FIFOs, dataflow tasks may form either acyclic or cyclic dependencies. In cyclic designs, an earlier-defined task in C may expect inputs from a later-defined task through a FIFO, creating a feedback loop. This significantly complicates simulation: sequential execution of C functions becomes invalid, as a function waiting for input from a subsequent module would stall indefinitely. Accurately simulating such systems requires modeling hardware concurrency using multithreading—a non-trivial task, as illustrated in our earlier example.

\section{Simulating Dataflow Designs at C Level}
\label{sec:simulate-dataflow}

\begin{figure*}[t]
    \centering
    \includegraphics[width=\linewidth]{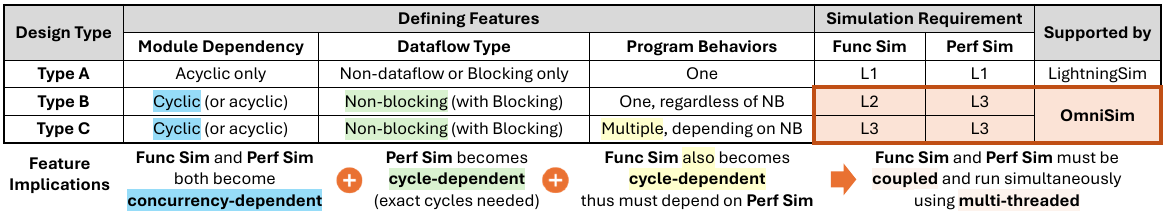}
    \vspace{-18pt}
    \caption{Defining features of three types of dataflow designs with implications on performance and functionality simulation.}
    \label{fig:type-table}
\end{figure*}

\begin{figure*}
    \centering
    \vspace{-4pt}
    \includegraphics[width=\linewidth]{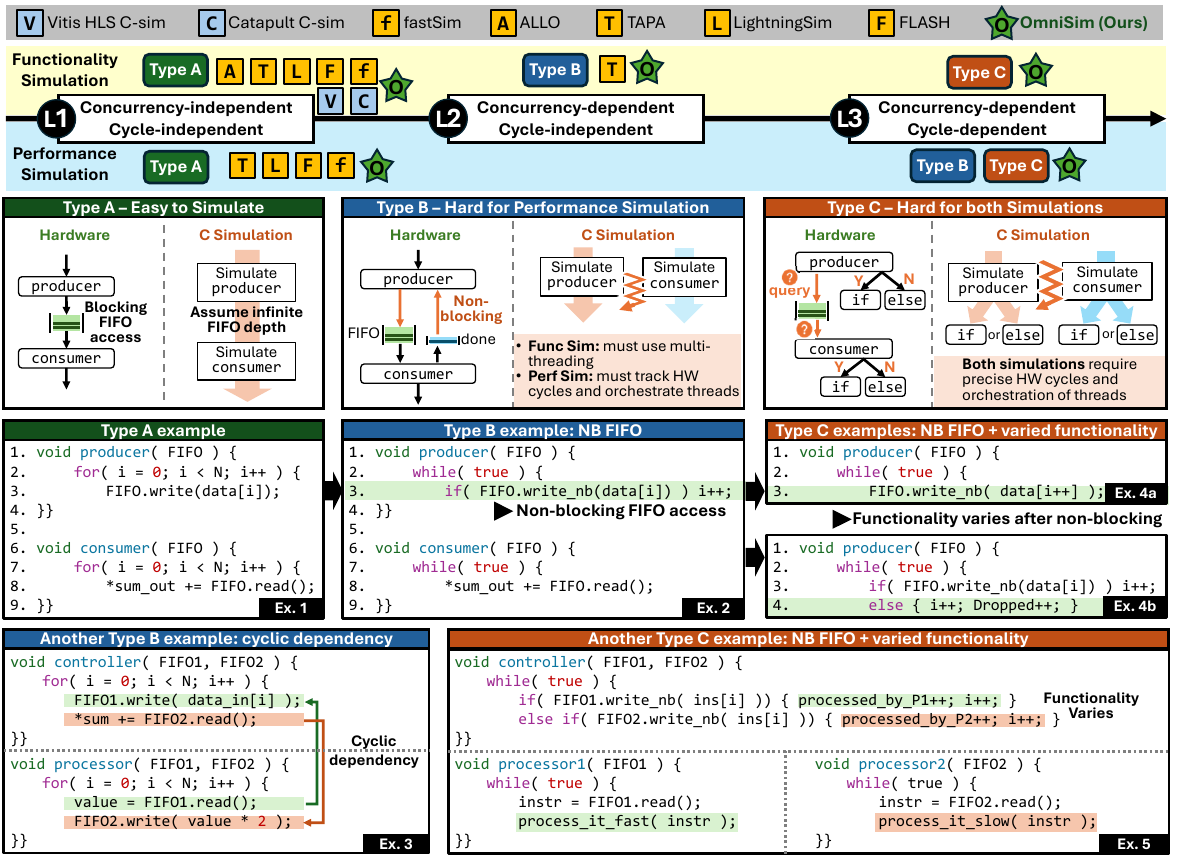}
    \vspace{-18pt}
    \caption{Taxonomy, classification, and examples of dataflow designs.}
    \label{fig:dataflow-types}
    \vspace{-8pt}
\end{figure*}

We introduce a detailed analysis and categorization of dataflow designs, highlighting their specific C-level simulation challenges. To the best of our knowledge, this is the first comprehensive study to propose a well-defined taxonomy and classification of such designs.

\vspace{-4pt}
\subsection{Taxonomy and Design Classification}
\label{sec:taxomony}

As shown in Fig.~\ref{fig:type-table}, we classify dataflow designs into three types based on three defining features: module dependency, FIFO access type, and the number of possible program behaviors after FIFO access. As will be detailed in Sec.~\ref{sec:implications}, each feature has specific implications for both functionality simulation (Func Sim) and performance simulation (Perf Sim), resulting in different levels of complexity. Supporting these defining features and addressing their simulation implications are the key motivations behind \thiswork{}, fundamentally distinguishing it from LightningSim.

\begin{itemize}[leftmargin=*]
  \item \textbf{Type A}: Designs that are either non-dataflow or use only \textit{blocking} FIFO accesses. Modules must have strictly \textit{acyclic} dependencies, and there is only \textit{one} possible program behavior after any FIFO access.
    \underline{Example.}  
  Fig.~\ref{fig:dataflow-types} Ex.~1 shows a basic setup where a \texttt{producer} sends data to a \texttt{consumer} via a blocking FIFO.
  \underline{Practical Usage.}  
  Type A is typical in fixed-function accelerators such as systolic arrays, convolution engines, and digital signal processing (DSP) pipelines, where data moves in a predictable, lockstep manner.

  \item \textbf{Type B}: Designs that may use \textit{non-blocking} FIFO accesses, include \textit{infinite loops}, or contain \textit{cyclic dependencies}, yet still exhibit only \textit{one} possible program behavior per FIFO access. That is, the success or failure of NB accesses \textit{does not alter} the program behavior.
  \underline{Examples.}  
  Ex.~2 extends Ex.~1 with NB FIFO accesses and infinite loops, modeling a polling-based stream processor.  
  Ex.~3 introduces a cyclic dependency between a \texttt{controller} and a \texttt{processor}, communicating via blocking FIFOs.
  \underline{Practical Usage.}  
  Type B designs are common to describe systems that respond to asynchronous or event-driven inputs without data-dependent branching. Examples include DMA engines polling memory-mapped FIFOs, stream filters that process packets as they arrive, or FSMs that iterate through actions. Cyclic dependencies are also common in systems with instructions and controllers, where feedback loops coordinate state updates and actuation.

  \item \textbf{Type C}: Designs that exhibit Type B characteristics but allow \textit{multiple possible behaviors} following a NB FIFO access. Such variability often arises from explicit control branches, data drops, or internal state updates in response to FIFO access outcomes.
    \underline{Examples.}  
  Ex.~4a and Ex.~4b modify Ex.~2 to introduce divergent behaviors when a write fails:  
  In Ex.~4a, data is silently dropped (\texttt{i++} still executes);  
  In Ex.~4b, an \texttt{if-else} branch tracks failures explicitly.  
  Ex.~5 shows a controller dispatching to one of two processors depending on which is less busy, dynamically responding to system state.
  \underline{Practical Usage.}  
Type C reflects the complexity of real-world hardware systems that adapt to dynamic resource, timing, or data conditions. Examples include out-of-order execution engines, adaptive packet routers, cache controllers handling hits/misses, dynamic memory allocators, and streaming accelerators that repartition workloads under backpressure.
\end{itemize}

\vspace{-4pt}
\subsection{Implications on Simulation Requirements}
\label{sec:implications}

The defining features of dataflow designs directly determine their simulation requirements. We identify two key considerations for accurate simulation:  
\textbf{(1) hardware concurrency (in)dependence} --- whether the simulation must model concurrent module execution (e.g., via multi-threading) to reflect simultaneous activity, and
\textbf{(2) hardware cycle (in)dependence} --- whether the simulation must capture the exact clock cycle at which hardware actions, such as FIFO reads and writes, occur.

Accordingly, there are three valid levels of simulation requirements, shown in the top row of Fig.~\ref{fig:dataflow-types}:  
\circled{L1} \emph{concurrency- and cycle-independent},  
\circled{L2} \emph{concurrency-dependent, cycle-independent}, and  
\circled{L3} \emph{concurrency- and cycle-dependent}.  
Only three combinations are possible, as cycle-dependence always implies concurrency-dependence.

The three defining features of design types map to these levels as annotated at the bottom of Fig.~\ref{fig:type-table}.  
First, designs with \emph{cyclic dependencies} require both functional and performance simulation to be concurrency-dependent, necessitating multi-threading.
Second, designs with \emph{NB FIFO accesses} require performance simulation to be cycle-dependent, where the timing of each FIFO access must be captured dynamically rather than determined statically.
Third, when the outcome of an NB access can change subsequent behavior, functional simulation also becomes cycle-dependent, because the functional correctness now depends on exact cycles.

These implications underscore the need for \textbf{tightly coupled functional and performance simulation executed concurrently via multi-threading}, which is the central goal of \thiswork{}.

\vspace{-2pt}
\subsubsection{Type A: Func Sim Level~1, Perf Sim Level~1}\mbox{}

\noindent Type~A designs can be simulated sequentially using a single thread for both Func Sim and Perf Sim.
Although the hardware executes modules concurrently, \textit{multi-threading is unnecessary} because FIFO access order is deterministic under the assumption of infinite FIFO depth.  
Consequently, both simulations are concurrency-independent and can be fully decoupled, as functional correctness does not depend on precise hardware cycles, and hardware cycle counts can be calculated statically after the fact based on Func Sim.

\vspace{-2pt}
\subsubsection{Type B: Func Sim Level~2, Perf Sim Level~3}\mbox{}

\para{Functionality simulation.}
Type~B designs are \textit{concurrency-dependent} for Func Sim, meaning multi-threading is required. This occurs when modules contain infinite loops that must execute in parallel.  
For example, in Ex.~2, if the \texttt{while} loop in the \texttt{producer} is simulated sequentially, any module defined after it (e.g., \texttt{consumer}) would never run.  
Similarly, in Ex.~3, which contains cyclic dependencies, the \texttt{controller} may wait for input from the \texttt{processor} before proceeding. Even without infinite loops, such dependencies require concurrent simulation.

However, Type~B designs are \textit{cycle-independent} for Func Sim, because the functionality correctness depends only on \emph{what} hardware actions occur (e.g., FIFO accesses), not on \emph{when} they occur.  
In Ex.~2, as long as the \texttt{producer}'s behavior after a successful NB write is correctly modeled, the exact cycle of the write, whether at cycle~5 or cycle~10, is irrelevant.  
In Ex.~3, although cyclic dependencies require multi-threading, blocking accesses ensure a single valid execution order, which can be enforced with multi-thread locks and thus no extra orchestration is needed.
In short, even if Perf Sim is inaccurate, Func Sim can still be correct for Type~B designs.

\para{Performance simulation.}
Unlike Func Sim, performance simulation of Type~B designs is \textit{cycle-dependent} due to NB FIFO accesses.
For example, in Ex.~2, suppose we measure the number of cycles required for the \texttt{producer} to perform five successful FIFO writes (break condition omitted).  
One OS thread simulates the \texttt{producer} and another simulates the \texttt{consumer}.  
If the OS scheduler happens to run the \texttt{producer} thread significantly earlier or more often than the \texttt{consumer} thread, the simulation may record extra failed write attempts before the consumer has time to read from the FIFO, leading to an overestimated cycle count.

\subsubsection{Type C: Func Sim Level~3, Perf Sim Level~3}\label{sec:type-c}\mbox{}

\noindent Type~C designs are concurrency- and cycle-dependent for both Func Sim and Perf Sim.
The simulator must capture not only \textit{what} actions occur but also \textit{when} they occur, since functional behavior can change depending on the outcome of a NB FIFO access.  
For example, in Fig.~\ref{fig:motivation}, the computed value of \texttt{cycles} may vary entirely due to OS-level thread scheduling rather than true hardware timing. 
Similarly, in Ex.~4 and Ex.~5, omitting the exact cycle of a FIFO access can lead to both functional errors and inaccurate performance estimates.  
Accurate simulation therefore requires precise modeling of \textit{actual hardware cycles} rather than relying on \textit{software-level timing} dictated by the OS, making Level~3 simulation fundamentally challenging.

\subsubsection{Additional Challenge: Deadlock Detection}\label{sec:type-b-deadlock-detection}\mbox{}

\noindent For Type~B and Type~C designs, a further challenge is detecting hardware-level deadlocks originated from flawed design logic.  
For example, if a producer waits for a consumer while the consumer simultaneously waits for the producer, the system can deadlock regardless of FIFO depth.  
A simulator must distinguish such deadlocks from temporary stalls and report them promptly to avoid indefinite hangs where all modules are idle.  
This differs from livelock, where modules remain active but make no forward progress; neither \thiswork{} nor C/RTL co-simulation detects livelocks.

\section{Prior Work}

Catapult and Vitis HLS are two representative commercial HLS tools, while LegUp~\cite{canis2011legup} and Bambu~\cite{ferrandi2021bambu} are open-source academic alternatives. Among recent academic efforts, TAPA~\cite{tapa} is a specialized tool designed to support multi-kernel dataflow designs using C++, and Allo~\cite{chen2024allo} is a Python-based HLS framework that generates Vitis-compatible C programs. 
In addition to these end-to-end HLS tools, several studies have focused on accelerating performance simulation at the C level, including FLASH~\cite{flash}, FastSim~\cite{fastsim}, HLPerf~\cite{hlperf}, and LightningSim~\cite{lightningsim,lightningsimv2}.

\para{Functionality simulation.}
All existing HLS tools, both commercial and academic, can simulate only certain types of dataflow designs at the C level, specifically, Type A in our taxonomy. As discussed in Sec.~\ref{sec:intro}, both Catapult and Vitis impose explicit restrictions on simulating non-blocking designs (Type B and C).
Among academic tools, TAPA can simulate Type A and B but not C, while Allo supports only Type A. For unsupported types, these tools either disallow or fall back to RTL simulation to ensure correctness.

\para{Performance simulation.}
To perform fast performance simulation,
FLASH uses C++ to model the timing of HLS designs by combining the original source code with scheduling information from the HLS tool.
FastSim models the generated Verilog using C and applies optimizations tailored to common HLS structures, such as FSMs.
HLPerf, designed for modeling GNN accelerators, prioritizes simulation speed over cycle accuracy by omitting full functional simulation. LightningSim integrates both functionality and timing simulation at the LLVM IR level by injecting hardware scheduling information into the IR.

For simple Type A designs, all of these tools demonstrate reasonably good performance and accuracy. Among them, LightningSim stands out as both the fastest and most accurate, achieving near-C execution speed with 99.9\% RTL-level accuracy. 
However, for more complex Type B and Type C designs, none of the previous tools are able to simulate correctly.

\begin{figure}[t]
    \centering
    \includegraphics[width=\linewidth]{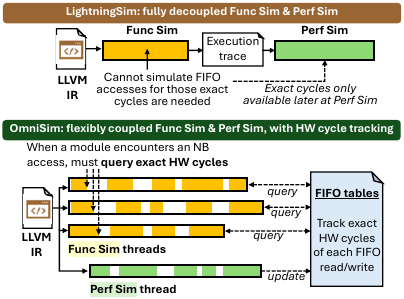}
    \caption{Comparing LightningSim and \thiswork{}.}
    \vspace{-8pt}
    \label{fig:new-Omni-vs-LS}
\end{figure}

\begin{figure*}
    \centering
    \includegraphics[width=\linewidth]{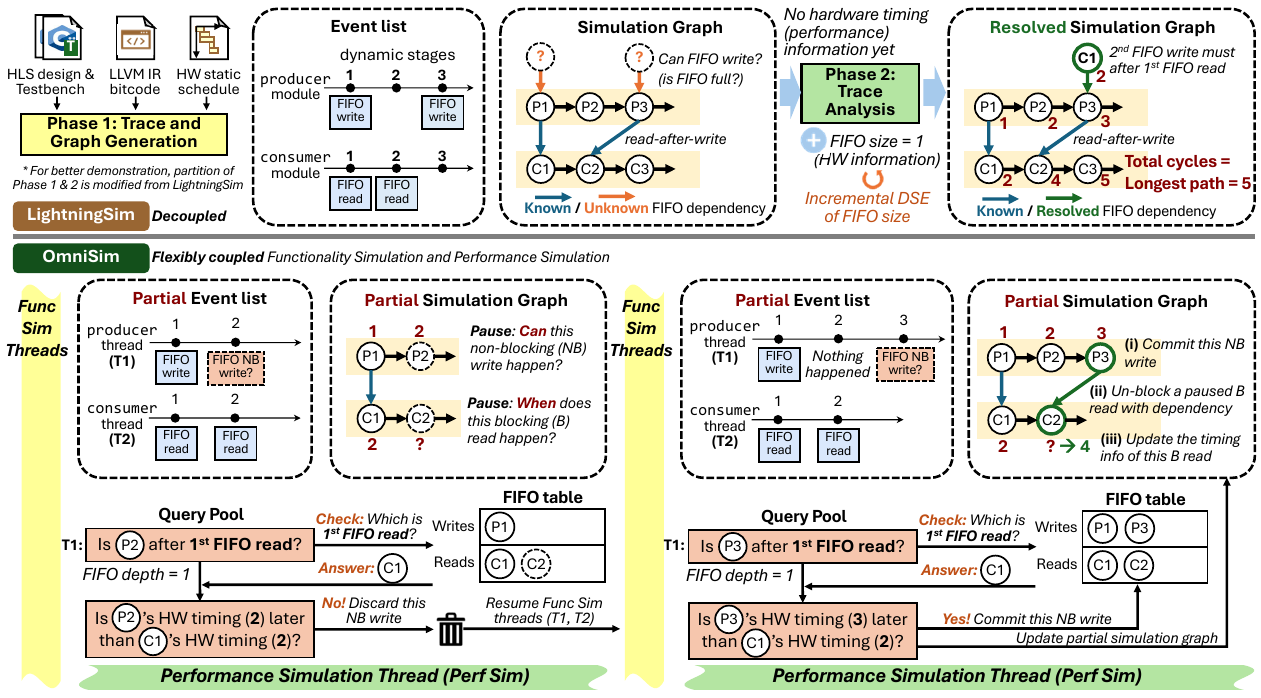}
    \vspace{-12pt}
    \caption{Overview of the \thiswork{} workflow compared to LightningSim. LightningSim executes functionality simulation and performance simulation as two fully decoupled phases. In contrast, \thiswork{} tightly integrates these two simulations, allowing flexible overlap to improve efficiency. The overlapping execution is illustrated in detail in Fig.~\ref{fig:Omni-threads}.}
    \label{fig:Omni-vs-LS}
\end{figure*}

\section{\thiswork{} Overview}
\label{sec:omni-overview}

Since \thiswork{} is inspired by LightningSim, we begin by comparing the two simulators and highlighting their key differences, followed by a simplified walk-through example that demonstrates the core techniques of \thiswork{}.

Fig.~\ref{fig:new-Omni-vs-LS} illustrates the key differences between LightningSim and \thiswork{}. LightningSim adopts a two-stage simulation model that fully decouples functionality simulation from performance simulation. This approach is effective for Type A designs, but is fundamentally limiting for Type B and C designs: as discussed in Sec.~\ref{sec:simulate-dataflow}, these design types require precise hardware timing information \textit{during} functionality simulation, which cannot be obtained in a decoupled simulation workflow.

In contrast, \thiswork{} introduces a fundamentally different simulation paradigm supported by novel data structures. First, it tightly couples functionality and performance simulation by coordinating multiple functionality simulation threads with a centralized performance simulation thread. The performance thread provides exact timing information to the functionality threads on demand.
Second, to ensure correct hardware timing regardless of arbitrary OS-level thread scheduling, \thiswork{} employs a robust thread orchestration mechanism synchronized via a set of FIFO timing tables. These tables dynamically track precise hardware cycle information for each FIFO access. Functionality threads query the tables as needed, while the performance thread updates them in real time.
Together, these two core techniques allow \thiswork{} to support all types of dataflow designs, making it significantly more general-purpose and capable than LightningSim.

\vspace{-4pt}
\subsection{LightningSim: Decoupled Two Phases}
\label{sec:bg-lightningsim}

Given a simple \texttt{producer}-\texttt{consumer} example with only blocking (B) FIFO accesses, LightningSim uses a single thread to simulate both modules sequentially.
As illustrated in Fig.~\ref{fig:Omni-vs-LS} (top), LightningSim operates in two distinct phases: \textit{untimed} functionality is simulated in Phase 1 through trace generation without hardware constraints, and \textit{timed} performance is simulated in Phase 2 via trace analysis.

\para{Phase 1: Trace and Simulation Graph Generation (untimed).}
Given an HLS design and testbench, LightningSim extracts the LLVM IR after front-end compilation,  instruments the IR for the host platform (for example, x86), and executes it to record a trace of the executed basic blocks. This phase performs functional simulation and generates the trace data required for performance modeling.

Next, it uses the \textit{static schedule} for each module—generated during C synthesis—to construct \textit{dynamic stages}. These stages are derived by ``unrolling'' the static schedule based on the execution trace. For example, if a loop has a static schedule of 1 and is executed four times with an initiation interval of one, the resulting dynamic stages will be labeled 1, 2, 3, and 4.

As shown in Fig.~\ref{fig:Omni-vs-LS}, the output of Phase 1 includes \textbf{event lists} and a \textbf{simulation graph}. An \textit{event} refers to an operation that may potentially cause a \textit{stall}, such as a FIFO access or a function call. Each module maintains an event list aligned with its dynamic stages, where specific stages are associated with events. A simulation graph is then constructed to capture dependencies among events, including function calls and FIFO read-after-write constraints. For instance, in the example shown, the \texttt{consumer}’s first read (C1) must follow the \texttt{producer}’s first write (P1), resulting in a directed edge from P1 to C1. Similarly, P3 precedes C2.

However, certain dependencies, such as FIFO write-after-read, remain unresolved at this stage. A module cannot write to a FIFO if it is full, but the simulator cannot determine fullness without knowing the FIFO’s capacity. These unresolved dependencies are marked as \textit{unknown} and are addressed in the next phase.

\para{Phase 2: Trace Analysis (timed with HW info).}
In this phase, LightningSim incorporates hardware-specific constraints, such as FIFO sizes, to perform \textit{stall analysis} and resolve the previously unknown dependencies.
In the same example, suppose the FIFO has a depth of one. The first write operation (P1) can proceed immediately, but the second write (P3) must wait until the first read (C1) completes. This dependency is enforced by adding an edge from C1 to P3, indicating that P3’s write must occur after C1’s read.
Finally, LightningSim computes the total execution latency by identifying the longest path from the start node to all other nodes in the simulation graph, resulting in a cycle-accurate performance simulation of the original HLS design.
In this example, the total cycle count is 5, where node C2 is stalled by one cycle from P3.

\vspace{-4pt}
\subsection{OmniSim: Interleaved Two Phases}
\label{sec:omnisim-example}

We now modify the example in Fig.~\ref{fig:Omni-vs-LS} (bottom) by introducing a non-blocking (NB) FIFO write at dynamic stage 2 of the \texttt{producer} module. This changes the design from Type A into Type B or C, requiring a fundamentally different simulation strategy. 

First, as discussed in Sec.~\ref{sec:simulate-dataflow}, multi-threading is now required to simulate both modules. \thiswork{} assigns one thread per dataflow module: thread \texttt{T1} simulates the \texttt{producer}, and thread \texttt{T2} simulates the \texttt{consumer}. These are referred to as \textbf{Func Sim threads}.

Second, the simulator must determine whether the FIFO is full precisely at the \textit{hardware cycle} when a NB FIFO write is issued (similarly for NB reads). This determination depends on how many reads from \texttt{T2} have occurred up to that cycle—not on the OS scheduling order of \texttt{T1} and \texttt{T2}. To resolve this, \thiswork{} introduces a dedicated \textbf{Perf Sim thread}, responsible for synchronizing NB accesses and maintaining FIFO status based on hardware timing.

\para{Thread orchestration and example walk-through.}
The threads must be carefully orchestrated.
\underline{First}, during simulation, each Func Sim thread proceeds independently, computes its own hardware timing (cycle count) accurately, just like LightningSim, until it encounters a NB FIFO event. In this example, \texttt{T1} and \texttt{T2} each complete one blocking FIFO access, resulting in nodes P1 and C1 in the \textbf{partial simulation graph}, with a dependency from P1 to C1. Accordingly, P1 is assigned cycle 1 and C1 is assigned cycle 2.

\underline{Second}, \texttt{T1} reaches a NB FIFO write at dynamic stage 2, denoted as P2, pauses, and issues a \textbf{query}: will this FIFO write succeed? This will be answered by the Perf Sim thread. Meanwhile, \texttt{T2} also pauses, waiting on a blocking FIFO read because data is unavailable.

\underline{Third}, once both Func Sim threads are paused, the Perf Sim thread becomes active to attempt to answer all queries. It maintains a \textbf{FIFO read/write table} to record all committed and pending accesses along with their associated hardware cycles. Instead of using a simple FIFO occupancy counter, this table is essential as it accurately tracks event timing despite the misalignment between software thread scheduling and actual hardware behavior.

In our example, assume the FIFO has a depth of one. The query from \texttt{T1}---``Can the NB write at cycle 2 succeed?''---is translated into: ``Has the first FIFO read completed before this write?'' If not, the FIFO is still full, and the write must fail. The Perf Sim thread consults the FIFO table, confirming that the first FIFO read occurred at node C1, which is also at cycle 2. The query is further turned into: ``Is P2 (cycle 2) strictly after C1 (cycle 2)?'' The answer is no. Therefore, the NB write at P2 fails and is discarded. No state change occurs at cycle 2, and the Func Sim threads resume execution.

\underline{Fourth}, at dynamic stage 3, \texttt{T1} attempts another NB FIFO write, labeled P3. The new query is: ``Is P3 (cycle 3) after C1 (cycle 2)?'' This time, the answer is yes. As a result:
(i) the NB write is committed by adding P3;
(ii) the pending blocking read at \texttt{T2} is unblocked by establishing a dependency from P3 to C2; and 
(iii) the simulation graph is updated with C2 assigned to cycle 4, indicating that the read is satisfied immediately following the successful write.

\underline{Finally}, the Func Sim and Perf Sim threads continue in this interleaved fashion until the simulation completes, yielding a final simulation graph with accurate dependencies and cycle counts.

\vspace{-4pt}
\section{\thiswork{} Implementation}

Building on the simplified overview of thread orchestration in Sec.~\ref{sec:omni-overview}, demonstrating its support for all types of dataflow designs, we now present the detailed implementation.
\thiswork{} consists of two stages, front-end compilation, and multi-thread execution.
In the second stage, we particularly focus on a key technique, where functionality and performance simulations are not only tightly, but also \textit{flexibly overlapped}, to improve the simulation efficiency.

\subsection{Front-end Compilation}\label{sec:frontend-compilation}

The compilation process in \thiswork{} requires three sets of inputs. From the user-provided HLS project, we extract (1) the LLVM IR bitcode generated by the HLS front-end compilation, and (2) the user’s testbench code, typically in C++, which invokes the HLS kernel with appropriate inputs and checks its outputs. Additionally, (3) we introduce our own runtime library, implemented as a Linux shared object (\texttt{.so}) file, which provides core simulation infrastructure such as FIFO buffers, AXI interfaces, and trace collection utilities. Anytime the HLS code invokes a hardware-level intrinsic such as FIFO accesses, it calls into this library, which passes FIFO data between modules and enables \thiswork{} to track them. However, just intercepting hardware intrinsics is not enough to correctly simulate all hardware behaviors---for instance, \thiswork{} must also make dataflow tasks run concurrently. In such cases, we must instead \textit{rewrite} the HLS user's code at the LLVM level.

\subsubsection{LLVM Passes}

The first step in the front-end is to apply a series of custom LLVM passes to the LLVM IR, preparing it for simulation. Similar to LightningSim, we insert calls to a tracing function at the beginning of each basic block to enable fine-grained tracking of runtime execution paths at the LLVM instruction level.
In addition, \thiswork{} introduces new LLVM passes to enhance simulation capability and performance, specifically, to rewrite dataflows to use threads, enabling tasks to execute concurrently as discussed in Sec.~\ref{sec:simulate-dataflow}. To do this, we implement an LLVM pass that identifies dataflow functions in the IR, extracts their constituent sub-tasks, and generates a dedicated wrapper function for each sub-task. %
Instead of invoking tasks directly, the original dataflow function is rewritten to launch a separate thread for each wrapper function and wait for all threads to complete before returning.

\subsubsection{Compiling and Linking}

After applying the custom LLVM passes, we compile the transformed LLVM IR into object code using \texttt{clang}. The testbench C/C++ code is also compiled at this stage. The resulting objects are linked together with our runtime shared library, which provides the runtime support for FIFO and AXI interface implementations and the logic for trace collection and analysis. The final output is a binary executable that runs the simulation.

\begin{figure*}
    \centering
    \includegraphics[width=\linewidth]{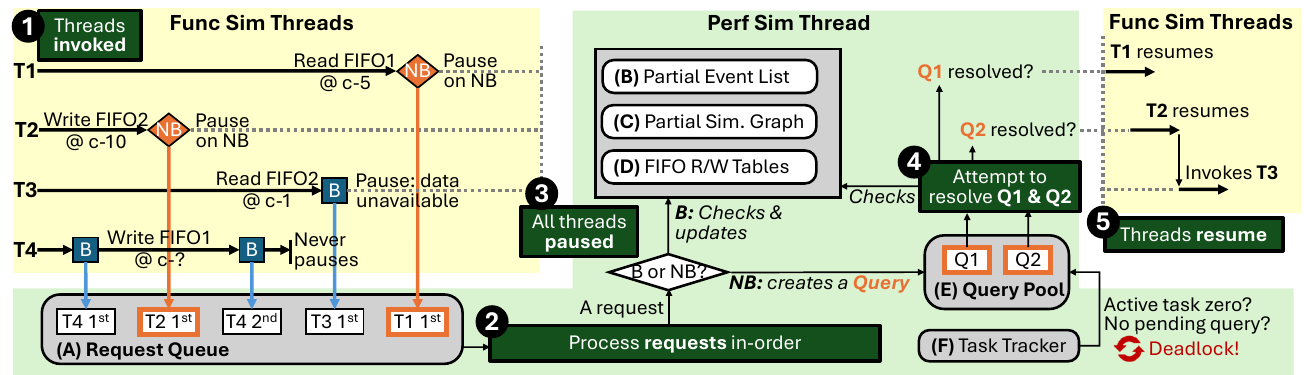}
    \caption{Execution workflow of \thiswork{}. Func Sim threads (yellow background) and the Perf Sim thread (green background) run concurrently with overlaps. Key steps include invoking a thread for each dataflow module, sending requests and pausing on queries, the Perf Sim thread processing requests and creating/resolving queries, and resuming Func Sim threads.}
    \label{fig:Omni-threads}
\end{figure*}

\subsection{Multi-thread Execution}
\label{sec:multi-thread-exec}

The execution of \thiswork{} consists of five major steps, annotated as \circled{1} to \circled{5} in Fig.~\ref{fig:Omni-threads}. Six key data structures are maintained during the simulation, labeled as \textbf{(A)} to \textbf{(F)}.

\vspace{2pt}
\noindent
\textbf{{\circled{1} Invoke all threads---Func Sim and Perf Sim.}}
At the beginning of execution, all necessary threads are launched, including one Perf Sim thread and multiple Func Sim threads. All dataflow modules, including submodules, will be allocated a dedicated thread; modules with only blocking accesses are also assigned a thread to support cyclic dependencies and infinite loops. In Fig.~\ref{fig:Omni-threads}, Func Sim threads are labeled \texttt{T1} through \texttt{T4}.

Once initiated, Func Sim threads begin simulating their respective modules and generating \textbf{requests}, which are inserted into a global \textbf{(A) Request Queue}. Table~\ref{tab:request_types} lists key (non-exhaustive) request types. Most are informative, such as blocking FIFO operations (e.g., \texttt{read()}, \texttt{write()}), and are used to update the simulation graph state, while NB accesses will result in \textbf{queries} that require resolution later, as listed in the last three rows of the table.
\begin{table}
    \centering
    \footnotesize
      \setlength\tabcolsep{2pt}
    \begin{tabular}{llc}
        \toprule
        \textbf{Request type} & \textbf{Description} & \textbf{Query?} \\
        \midrule
        \texttt{TraceBlock} & A basic block was executed & \emptycirc \\
        \texttt{StartTask} & A dataflow task started in a new thread & \emptycirc \\
        \texttt{FifoRead}/\texttt{Write} & FIFO was read from/written to & \emptycirc \\
        \texttt{AxiReadReq}/\texttt{WriteReq} & A read/write request issued on AXI & \emptycirc \\
        \texttt{AxiRead}/\texttt{Write} & AXI was read from/written to & \emptycirc \\
        \texttt{AxiWriteResp} & A write response was issued on AXI & \emptycirc \\
        \texttt{FifoCanRead}/\texttt{Write} & Query for FIFO empty/full & \fullcirc \\
        \texttt{FifoNbRead} & An NB FIFO read attemptted & \fullcirc \\
        \texttt{FifoNbWrite} & An NB FIFO write attemptted & \fullcirc \\
        \bottomrule
    \end{tabular}
    \caption{Key requests made by Func Sim threads.}
    \vspace{-16pt}
    \label{tab:request_types}
\end{table}

\vspace{2pt}
\noindent
\textbf{{\circled{2} Process all requests---by Perf Sim thread.}}
While Func Sim threads are running, the Perf Sim thread operates concurrently, actively processing requests from the queue.
As illustrated in Fig.~\ref{fig:Omni-threads}, yellow highlights indicate active periods of Func Sim threads, and green highlights indicate the active periods of the Perf Sim thread. Notably, the overlap between these periods enables \thiswork{} to not only support Type B and C dataflow designs but also to simulate Type A designs more efficiently than LightningSim, as will be demonstrated in the evaluation.

As soon as Func Sim threads begin inserting requests into the request queue, the Perf Sim thread starts processing them. It checks and updates key data structures: \textbf{(B)} partial event list, \textbf{(C)} partial simulation graph, and \textbf{(D)} FIFO read/write tables. In addition, for a request that is also a NB access, the Perf Sim thread creates a \textbf{query} and places it into the \textbf{(E) Query Pool} for resolution.

\vspace{2pt}
\noindent
\textbf{{\circled{3} All Func Sim threads pause.}}
A Func Sim thread pauses whenever it encounters a query that must be answered, such as a NB read/write or FIFO status check. For example, \texttt{T1} pauses at a NB read from FIFO 1 at hardware cycle c-5, since it needs to determine whether the FIFO is non-empty at that moment. Likewise, \texttt{T2} pauses on a NB write to FIFO 2 at c-10 to determine whether the FIFO is full.
This example again illustrates that hardware timing and software thread execution order may not align—e.g., \texttt{T1} simulating cycle c-5 may be scheduled after \texttt{T2} simulating cycle c-10.

Func Sim threads may also pause on blocking reads if the corresponding FIFO is empty. In the example, \texttt{T3} pauses on a blocking read from FIFO 2, which will not have data until \texttt{T2} writes to it at cycle c-10.
An exception is a thread that performs only blocking writes, such as \texttt{T4}. These threads assume infinite FIFO depth and do not pause, but they still contribute events to the simulation.

Each time a Func Sim thread is activated, the \textbf{(F) Task Tracker} is incremented, and when a thread pauses, it is decremented. When the Task Tracker reaches zero (i.e., all Func Sim threads are paused), the Perf Sim thread begins resolving queries.

\vspace{2pt}
\noindent
\textbf{{\circled{4} Attempt to resolve queries---by the Perf Sim thread.}}
All queries in the query pool are examined by the Perf Sim thread. For each query, the thread checks the FIFO read/write table to determine whether the corresponding NB access can proceed. Unlike LightningSim, \thiswork{} must consider hardware constraints, particularly FIFO sizes, to process NB FIFO accesses.

Each query has a \textit{source} and a \textit{target}: The \textit{source} is the NB event being queried (e.g., the $w^{\text{th}}$ write or $r^{\text{th}}$ read), and the \textit{target} is the committed FIFO access against which the source must be compared. Let the FIFO size be $S$; the resolution logic is shown in Table~\ref{tab:query-resolution}.
Specifically, if $w \leq S$, then the NB write is guaranteed to succeed. If $w > S$, the thread checks whether the $(w - S)^{\text{th}}$ read has already occurred. Similarly, an $r^{\text{th}}$ NB read is allowed only if the $r^{\text{th}}$ write has already completed.

If the target event has already been simulated, the query can be resolved by comparing the actual hardware cycle values, as described in Sec.~\ref{sec:omni-overview}, which are maintained using the partial simulation graph. If the target has not yet been simulated, i.e., being \texttt{unknown}, the query remains in the pool and will be revisited in the next Perf Sim iteration.
This is possible because of arbitrary OS scheduling.

\begin{table}[t]
\centering
\footnotesize
\begin{tabular}{c|c|c|c}
\toprule
\textbf{Query Type} & \textbf{Source} & \textbf{Target} & \textbf{Resolution Condition} \\ \midrule
NB Write & $w^{\text{th}}$ write & N/A & \texttt{true} if $w \leq S$ \\ 
NB Write & $w^{\text{th}}$ write & $(w - S)^{\text{th}}$ read & \texttt{true} if source is after target \\ 
NB Read  & $r^{\text{th}}$ read  & $r^{\text{th}}$ write       & \texttt{true} if source is after target \\ \bottomrule
\end{tabular}
\caption{Resolution for NB queries, assuming FIFO size $S$.}
\label{tab:query-resolution}
\vspace{-24pt}
\end{table}

\vspace{2pt}
\noindent
\textbf{{\circled{5} Func Sim threads resume.}}
Each time a query is resolved, the result is sent to the waiting Func Sim thread, which resumes execution immediately. The task tracker is updated accordingly.
Some queries also have follow-up effects, depending on the result. Specifically, if a NB read succeeds, the FIFO is read immediately and the simulation graph is updated.
If a NB write succeeds, it may unblock a pending blocking read in another thread.
For example, in Fig.~\ref{fig:Omni-threads}, a blocking read by \texttt{T3} is automatically unpaused after a successful NB write by \texttt{T2}.

\vspace{2pt}
\noindent
\textbf{{Finalization.}}
Once all Func Sim threads finish execution, simulation is complete. The final simulation graph is now fully constructed. We traverse the entire graph one more time to compute the total simulation cycle count, including accurate timing for previous blocking writes and other delayed events.

\section{\thiswork{} Optimizations} 

In addition to the flexibly coupled Func Sim and Perf Sim technique introduced in Sec.~\ref{sec:multi-thread-exec}, \thiswork{} incorporates several additional optimizations that are either essential to its core simulation capabilities or critical to its efficiency.

\subsection{Deadlock Detection}
A critical requirement to simulate dataflow designs is to correctly detect deadlocks, which are especially common due to cyclic dependencies and improperly sized FIFOs.

\underline{First}, a fundamental challenge shared by all simulators is distinguishing true, design-induced deadlocks from temporary stalls, without causing the simulator itself to hang infinitely. To address this, \thiswork{} employs the \textbf{(F) Task Tracker}, described in Sec.~\ref{sec:multi-thread-exec}, which tracks the number of active Func Sim threads currently executing HLS code, i.e., those not waiting for a FIFO read or write. When this counter reaches zero, the Perf Sim thread checks for unresolved queries. If no queries remain, it indicates that all Func Sim threads are blocked—either on empty FIFO reads or full FIFO writes. In such cases, \thiswork{} immediately detects and reports a true design-level deadlock.

\underline{Second}, \thiswork{} faces another challenge: when the Perf Sim thread is unable to resolve any pending queries because all target events remain \texttt{unknown}, simply returning control to the Func Sim threads would deadlock the simulator itself, as no thread can proceed. To handle this, \thiswork{} leverages a key insight: this situation implies that all threads have progressed to at least the clock cycle of the earliest unresolved query.\footnote{Some threads may be stalled on FIFO reads rather than queries. However, since no other task can provide writes to unblock these reads, they must remain paused until at least the time of the earliest query.}
Since the target event of the earliest query is still \texttt{unknown}, it must lie in the future relative to the query. Therefore, that query can be safely resolved as \texttt{false} (e.g., the target FIFO read has not yet occurred, so the NB write fails; similar for vice versa). By resolving the earliest query, it enables at least one Func Sim thread to resume execution, ensuring forward progress and avoiding simulator-level deadlock.

\subsection{Incremental Simulation}\label{sec:incremental-simulation}

Optimizing FIFO sizes in dataflow designs is crucial for both performance and hardware resource: smaller FIFOs may introduce stalls or even deadlocks, while larger FIFOs consume more area. In traditional HLS flows, optimizing FIFO sizes is notoriously difficult—every modification requires re-running slow C/RTL co-sim to accurately assess the performance impact.

A key advantage of LightningSim is its support for highly efficient incremental simulation when FIFO sizes change, as illustrated in Fig.~\ref{fig:Omni-vs-LS} (top). In such cases, LightningSim can update performance estimates in just a few milliseconds during Phase 2. This is enabled by its decoupled functionality and performance simulations: the same simulation graph is reused, and only the longest-path analysis is rerun, without reconstructing the graph. For Type A designs, where functional behavior is independent of FIFO depths, the simulation graph remains valid across all FIFO configurations.

However, \thiswork{} builds its simulation graph dynamically during execution, based on the specific FIFO sizes used. This raises a key question: given a new set of FIFO sizes, can the existing simulation graph be reused, or is a full re-simulation required?

\label{sec:constraints}To address this, we introduce the concept of \textbf{constraints}. After each query is resolved as either \texttt{true} or \texttt{false}, \thiswork{} records the outcome as a constraint associated with the original query in the simulation graph. When a new FIFO configuration is provided, \thiswork{} validates the feasibility of incremental simulation by re-running the \textbf{Finalization} step (Sec.~\ref{sec:multi-thread-exec}) under the new depths to compute updated cycle counts. Each stored constraint is then re-evaluated against the new results.
If any query now resolves differently, for example, a NB access that previously succeeded would now fail, it indicates that the control or data flow may diverge from the original execution. In such cases, the existing simulation graph is no longer valid, and a full re-simulation is required. Otherwise, the graph can be reused to incrementally compute new cycle counts.

\subsection{Runtime Optimizations}

\subsubsection{Graph structure.}
In LightningSimV2, the simulation graph is generated and stored in the compressed sparse row (CSR) format, under the assumption that it will not be traversed until fully constructed. While CSR is efficient for traversal, it relies on complex intermediate data structures to hold node and event information that has not yet been committed to the graph.

However, \thiswork{} requires frequent traversal of the partial simulation graph to resolve queries. To avoid copying the graph and committing all intermediate data at each evaluation, which is prohibitively expensive, \thiswork{} adopts an adjacency list representation for efficient construction and traversal. We specialize the format to store one edge alongside each node, minimizing pointer chasing. We also prune unnecessary nodes to reduce overhead.
This design ensures that \thiswork{} can perform zero-copy traversal of the incomplete graph, without requiring temporary modifications.%

\subsubsection{Eliminating redundant FIFO checks.}
Evaluating non-blocking FIFO checks like \texttt{empty()} and \texttt{full()} can incur unnecessary runtime overhead, especially if their return values are never used. To address this, \thiswork{} introduces an LLVM pass to identify calls to such functions where the result is unused, replacing them with special markers indicating that the check can safely be skipped.

\section{Experiments}
\label{sec:result}

\begin{table*}
    \centering
    \footnotesize
    \renewcommand{\arraystretch}{1.05}
        \setlength\tabcolsep{2pt}
    \begin{tabular}{r|p{0.3\linewidth}|p{0.3\linewidth}|p{0.3\linewidth}|}
        \multicolumn{1}{r}{} & \multicolumn{1}{c}{\normalsize\textbf{C-sim}} & \multicolumn{1}{c}{\normalsize\textbf{Co-sim}} & \multicolumn{1}{c}{\normalsize\textbf{\thiswork}} \\\cline{2-4}
        {\texttt{fig\ref{fig:dataflow-types}\_ex2}} & \texttt{\textcolor{red}{@E Simulation failed: SIGSEGV.}} & {\footnotesize\texttt{sum\_out = \textcolor{Green}{2051325}}} & {\footnotesize\texttt{sum\_out = \textcolor{Green}{2051325}}} \\\cline{2-4}
        {\texttt{fig\ref{fig:dataflow-types}\_ex3}} & \texttt{\textcolor{DarkGoldenrod}{WARNING1}\textrm{\enspace\textcolor{gray}{(\texttimes{}2025)}}; \textcolor{DarkGoldenrod}{WARNING2}; sum = \textcolor{red}{0}} & \texttt{sum = \textcolor{Green}{4098600}} & \texttt{sum = \textcolor{Green}{4098600}} \\\cline{2-4}
        {\texttt{fig\ref{fig:dataflow-types}\_ex4a}} & \texttt{sum\_out = \textcolor{red}{2051325}} & \texttt{sum\_out = \textcolor{Green}{684453}} & \texttt{sum\_out = \textcolor{Green}{684453}} \\\cline{2-4}
        {\texttt{fig\ref{fig:dataflow-types}\_ex4a\_d}} & \texttt{\textcolor{red}{@E Simulation failed: SIGSEGV.}} & \texttt{sum\_out = \textcolor{Green}{684453}} & \texttt{sum\_out = \textcolor{Green}{684453}} \\\cline{2-4}
        {\texttt{fig\ref{fig:dataflow-types}\_ex4b}} & \texttt{sum\_out = \textcolor{red}{2051325}; Dropped = \textcolor{red}{0}} & \texttt{sum\_out = \textcolor{Green}{684453}; Dropped = \textcolor{Green}{1348}} & \texttt{sum\_out = \textcolor{Green}{684453}; Dropped = \textcolor{Green}{1348}} \\\cline{2-4}
        {\texttt{fig\ref{fig:dataflow-types}\_ex4b\_d}} & \texttt{\textcolor{red}{@E Simulation failed: SIGSEGV.}} & \texttt{sum\_out = \textcolor{Green}{684453}; Dropped = \textcolor{Green}{1358}} & \texttt{sum\_out = \textcolor{Green}{684453}; Dropped = \textcolor{Green}{1358}} \\\cline{2-4}
        {\texttt{fig\ref{fig:dataflow-types}\_ex5}} & \texttt{processed\_by\_P1 = \textcolor{red}{2025}; processed\_by\_P2 = \textcolor{red}{0}\newline{}sum\_out\_P1 = \textcolor{red}{2051325}; sum\_out\_P2 = \textcolor{red}{0}} & \texttt{processed\_by\_P1 = \textcolor{Green}{1351}; processed\_by\_P2 = \textcolor{Green}{674}\newline{}sum\_out\_P1 = \textcolor{Green}{1366878}; sum\_out\_P2 = \textcolor{Green}{684447}} & \texttt{processed\_by\_P1 = \textcolor{Green}{1351}; processed\_by\_P2 = \textcolor{Green}{674}\newline{}sum\_out\_P1 = \textcolor{Green}{1366878}; sum\_out\_P2 = \textcolor{Green}{684447}} \\\cline{2-4}
        {\texttt{fig\ref{fig:motivation}\_timer}} & \texttt{\textcolor{DarkGoldenrod}{WARNING1}\textrm{\enspace\textcolor{gray}{(\texttimes{}2025)}}; \textcolor{DarkGoldenrod}{WARNING2}\newline{}Internal timer counted \textcolor{red}{0} cycles} & \texttt{Internal timer counted \textcolor{Green}{6075} cycles} & \texttt{Internal timer counted \textcolor{Green}{6075} cycles} \\\cline{2-4}
        {\texttt{deadlock}} & \texttt{\textcolor{DarkGoldenrod}{WARNING1}\textrm{\enspace\textcolor{gray}{(\texttimes{}2025)}}; \textcolor{DarkGoldenrod}{WARNING2}; \textcolor{red}{sum = 0}} & \texttt{ERROR!!! \textcolor{Green}{DEADLOCK DETECTED} at 5145000 ns! SIMULATION WILL BE STOPPED!} & \texttt{thread 'omnisim' panicked at \textrm{[\ldots]}:\newline{}\textcolor{Green}{unresolvable deadlock detected}} \\\cline{2-4}
        {\texttt{branch}} & \texttt{\textcolor{DarkGoldenrod}{WARNING2}; fetched = \textcolor{red}{2025}; executed = \textcolor{red}{102}} & \texttt{fetched = \textcolor{Green}{955}; executed = \textcolor{Green}{51}} & \texttt{fetched = \textcolor{Green}{955}; executed = \textcolor{Green}{51}} \\\cline{2-4}
        {\texttt{multicore}} & \texttt{\textcolor{DarkGoldenrod}{WARNING2}\textrm{\enspace\textcolor{gray}{(\texttimes{}16)}}\newline{}total\_fetched = \textcolor{red}{32400}; total\_executed = \textcolor{red}{1221}} & \texttt{total\_fetched = \textcolor{Green}{15519}\newline{}total\_executed = \textcolor{Green}{614}} & \texttt{total\_fetched = \textcolor{Green}{15519}\newline{}total\_executed = \textcolor{Green}{614}} \\\cline{2-4}
        \multicolumn{4}{c}{\texttt{WARNING1: Hls::stream \textrm{[\ldots]} is read while empty, \textrm{[\ldots]}} \qquad \texttt{WARNING2: Hls::stream \textrm{[\ldots]} contains leftover data, \textrm{[\ldots]}}}
    \end{tabular}
    \vspace{2pt}
    \caption{Comparison of Func Sim across C-sim, Co-sim, and OmniSim for Type B and Type C designs. C-sim fails to correctly simulate the functionality of these designs, whereas OmniSim matches Co-sim exactly.}
    \vspace{-8pt}
    \label{fig:outputs}
\end{table*}

\begin{table}
    \centering
    \footnotesize
    \setlength\tabcolsep{2pt}
    \begin{tabular}{lcccccl}
        \toprule
        \textbf{Name} & \textbf{Type} & \textbf{\#Mod} & \textbf{\#FIFO} & \textbf{B/NB} & \textbf{Cyclic?} & \textbf{Description} \\
        \midrule
        \texttt{fig\ref{fig:dataflow-types}\_ex2} & B & 3 & 2 & NB & Yes & NB FIFO access (done signal) \\
        \texttt{fig\ref{fig:dataflow-types}\_ex3} & B & 3 & 2 & B & Yes & Cyclic dependency \\
        \texttt{fig\ref{fig:dataflow-types}\_ex4a} & C & 3 & 1 & NB & No & Skip if FIFO full \\
        \texttt{fig\ref{fig:dataflow-types}\_ex4a\_d} & C & 3 & 2 & NB & Yes & Skip if full (done signal) \\
        \texttt{fig\ref{fig:dataflow-types}\_ex4b} & C & 3 & 1 & NB & No & Count dropped elements \\
        \texttt{fig\ref{fig:dataflow-types}\_ex4b\_d} & C & 3 & 2 & NB & Yes & Count dropped (done signal) \\
        \texttt{fig\ref{fig:dataflow-types}\_ex5} & C & 4 & 2 & NB & No & Congestion-aware select \\
        \texttt{fig\ref{fig:motivation}\_timer} & C & 3 & 2 & NB & Yes & Fixed-point cycle count \\
        \texttt{deadlock} & B & 3 & 2 & B & Yes & Mutual blocking read \\
        \texttt{branch} & C & 3 & 2 & NB & Yes & Branch instructions \\
        \texttt{multicore} & C & 34 & 64 & NB & Yes & Multiple cores with branches \\
        \bottomrule
    \end{tabular}
    \caption{Evaluated Type B and Type C designs.}
    \vspace{-16pt}
    \label{tab:dataset}
\end{table}

In this section, we evaluate the two advantages of \thiswork{}: its ability to extend C-level HLS simulation to support complex dataflow designs, and its significantly improved simulation speed over state-of-the-art simulators.

First, we develop a set of \textbf{Type B} and \textbf{Type C} designs using Vitis HLS—designs that existing tools fail to simulate accurately at the C level (Sec.~\ref{sec:type-c})—and show that \thiswork{} can correctly perform both functionality and performance simulations.

Second, we evaluate the speed of \thiswork{}, comparing it against co-simulation for the Type B and C designs, as well as \textit{directly against} the state-of-the-art HLS simulator LightningSimV2~\cite{lightningsimv2} on its own benchmark suite, which consists entirely of Type A designs. %

All experiments are conducted on a server with a 64-core Intel Xeon Gold 6226R x86-64 CPU and 502 GiB of RAM, running Red Hat Enterprise Linux Server 8.10 and Xilinx Vitis HLS 2021.1.

\subsection{Simulating Complicated Dataflow Designs}\label{sec:results-making-it-possible}

\subsubsection{Design Dataset}

Because of the limitations of HLS in simulating Type B and Type C designs, unfortunately, designers tend not to use HLS in these cases, making it hard to find real-world open-source examples of such designs. To the best of our knowledge, no existing HLS benchmark suite includes \textit{any} Type B or Type C designs. Our dataset of Type B and C designs therefore consists mostly of smaller examples; however, these examples are only used for demonstration and do not limit \thiswork{}'s generality.

We enumerate these designs in Table~\ref{tab:dataset}.\footnote{Designs available at \url{https://github.com/sharc-lab/omnisim-benchmarks}.}
First, we include all the Type B and Type C designs shown in Fig.~\ref{fig:dataflow-types}, including non-blocking accesses in an infinite loop (terminated by a done signal), cyclic dependencies, and cycle-dependent functionality variance. We also implement the \texttt{timer} pattern shown in Fig.~\ref{fig:motivation}, another Type C design, to test \thiswork{}'s ability to orchestrate threads independent of OS thread scheduling. We test \thiswork{}'s deadlock detection capability with a cyclic dataflow designed to deadlock, in which two tasks block trying to read empty FIFOs. We also include a design called \texttt{branch} where a downstream executor task influences an upstream data fetcher, and a larger \texttt{multicore} version.

\subsubsection{Functionality Simulation}

\begin{figure}
\vspace{-12pt}
    \centering
    \includegraphics[width=\linewidth]{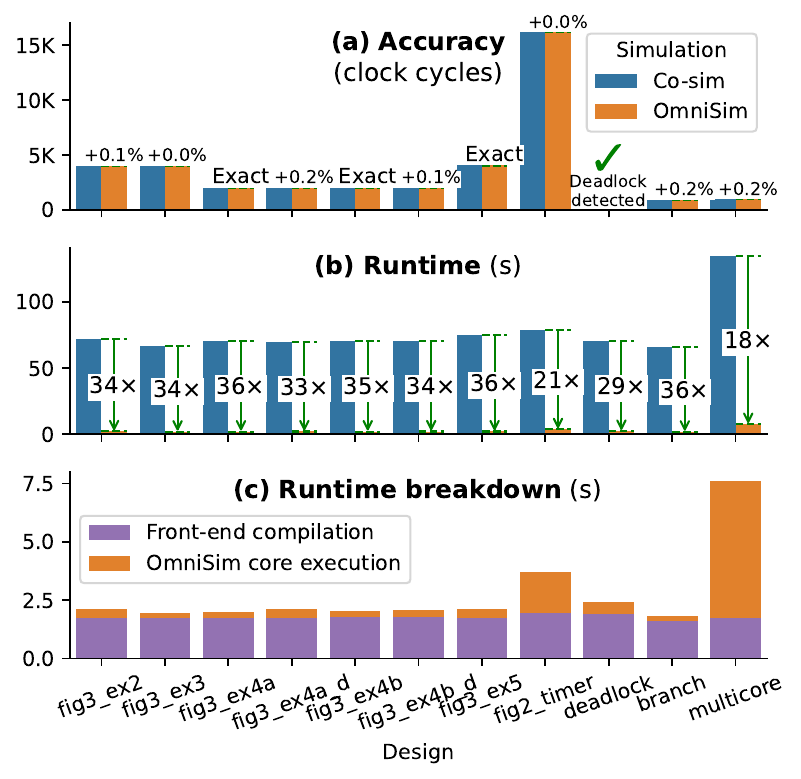}
    \vspace{-22pt}
    \caption{Comparisons of \thiswork's (a) cycle-accuracy and (b) simulation runtime against co-sim. A more detailed breakdown of the \thiswork{} execution time is shown in (c).}%
    \vspace{-10pt}
    \label{fig:compare-cosim}
\end{figure}

We compare the outputs of each design using different simulation methods in Table~\ref{fig:outputs}. Clearly, \textbf{\textit{none can be correctly simulated by C-sim}}. For instance, three of our designs (\texttt{fig\ref{fig:dataflow-types}\_ex2}, \texttt{fig\ref{fig:dataflow-types}\_ex4a\_d}, and \texttt{fig\ref{fig:dataflow-types}\_ex4b\_d}) contain infinite loops reading the input array in the \texttt{producer} task that are terminated by a done signal from the \texttt{consumer} task. In C-sim, \textit{these designs crash} because they run off the end of the input array but never proceed to simulate the \texttt{consumer} task that sends the done signal.
Even designs that do not fail so catastrophically still exhibit strange and inaccurate behavior under C-sim. For instance, \texttt{fig\ref{fig:dataflow-types}\_ex3} emits many warnings indicating that a stream ``is read while empty'' or ``contains leftover data'', and outputs zero in place of the expected sum. Meanwhile, other designs---like \texttt{fig\ref{fig:dataflow-types}\_ex4a}, \texttt{fig\ref{fig:dataflow-types}\_ex4b}, and \texttt{fig\ref{fig:dataflow-types}\_ex5}---silently give incorrect results based on the wrong assumption that \texttt{write\_nb()} calls always succeed.

In contrast, the output of each design under \thiswork{} matches the expected output produced by co-sim in every case---including producing the correct timer output in \texttt{fig\ref{fig:motivation}\_timer} (6075 cycles), as well as detecting the deadlock in the \texttt{deadlock} design.

\subsubsection{Performance Simulation}

\thiswork{} also obtains highly accurate cycle counts for each of these hard-to-simulate designs, closely aligning with the cycle counts produced by C/RTL co-simulation. Results are shown in Fig.~\ref{fig:compare-cosim}(a). We observe cycle counts that are, on average, only \textbf{0.09\% off} from the cycle counts reported by co-simulation. Even though LightningSim cannot support these designs, we benefit from the high accuracy of LightningSim's simulation graph approach atop which \thiswork{} was developed.

\subsection{Simulation Speed}

\subsubsection{Type B and C Designs}

Fig.~\ref{fig:compare-cosim}(b) compares \thiswork{}  speed against co-simulation, achieving a geomean of \textbf{30.7\texttimes{} speedup}. In the best case, achieving 35.9\texttimes{} speedup, \thiswork{} reduces the simulation from \textbf{66 seconds} by co-sim to only \textbf{1.83 seconds} at C level.
We also emphasize that our reported times are \textit{end-to-end}, including the time needed for front-end compilation (Sec.~\ref{sec:frontend-compilation}), which, in fact, usually dominates the majority of the \thiswork{} runtime (shown in purple in Fig.~\ref{fig:compare-cosim}(c)).

\subsubsection{Versus LightningSimV2}

\begin{table}
    \label{tab:compare-lsv2}
    \centering
    \setlength{\tabcolsep}{2.5pt}
    \small
    \begin{tabular}{l|c|ccc|c}
        \toprule
        & \textbf{LSv2} & \multicolumn{3}{c|}{\textbf{OmniSim}} & \\
        \textbf{Benchmark} & \textbf{Total} & \textbf{Total} & \textbf{FE} & \textbf{MT} & $\mathbf{\Delta\downarrow}$ \\
        \midrule
        Fixed-point square root~\cite{xilinx2021basic} & 4.97 & 3.65 & 3.42 & 0.23 & 1.36\texttimes{} \\
        FIR filter~\cite{xilinx2021basic} & 2.43 & 1.94 & 1.74 & 0.20 & 1.25\texttimes{} \\
        Fixed-point window conv~\cite{xilinx2021basic} & 3.69 & 3.15 & 2.77 & 0.38 & 1.17\texttimes{} \\
        Floating point conv~\cite{xilinx2021basic} & 2.42 & 2.46 & 2.11 & 0.35 & 0.98\texttimes{} \\
        Arbitrary precision ALU~\cite{xilinx2021basic} & 2.12 & 2.03 & 1.81 & 0.22 & 1.04\texttimes{} \\
        Parallel loops~\cite{xilinx2021basic} & 2.34 & 2.16 & 1.94 & 0.23 & 1.08\texttimes{} \\
        Imperfect loops~\cite{xilinx2021basic} & 2.24 & 2.13 & 1.91 & 0.23 & 1.05\texttimes{} \\
        Loop with max bound~\cite{xilinx2021basic} & 2.25 & 2.14 & 1.91 & 0.23 & 1.05\texttimes{} \\
        Perfect nested loops~\cite{xilinx2021basic} & 2.27 & 2.12 & 1.90 & 0.22 & 1.07\texttimes{} \\
        Pipelined nested loops~\cite{xilinx2021basic} & 2.23 & 2.19 & 1.96 & 0.23 & 1.02\texttimes{} \\
        Sequential accumulators~\cite{xilinx2021basic} & 2.29 & 2.20 & 1.98 & 0.22 & 1.04\texttimes{} \\
        Accumulators + asserts~\cite{xilinx2021basic} & 2.30 & 2.30 & 2.08 & 0.22 & 1.00\texttimes{} \\
        Accumulators + dataflow~\cite{xilinx2021basic} & 2.29 & 2.19 & 1.95 & 0.24 & 1.05\texttimes{} \\
        Static memory example~\cite{xilinx2021basic} & 2.18 & 2.12 & 1.92 & 0.19 & 1.03\texttimes{} \\
        Pointer casting example~\cite{xilinx2021basic} & 2.15 & 2.13 & 1.94 & 0.19 & 1.01\texttimes{} \\
        Double pointer example~\cite{xilinx2021basic} & 2.14 & 1.91 & 1.71 & 0.19 & 1.12\texttimes{} \\
        AXI4 master~\cite{xilinx2021basic} & 2.19 & 2.07 & 1.90 & 0.17 & 1.06\texttimes{} \\
        AXIS w/o side channel~\cite{xilinx2021basic} & 2.06 & 1.94 & 1.77 & 0.17 & 1.06\texttimes{} \\
        Multiple array access~\cite{xilinx2021basic} & 2.18 & 2.08 & 1.87 & 0.21 & 1.05\texttimes{} \\
        Resolved array access~\cite{xilinx2021basic} & 2.20 & 2.05 & 1.84 & 0.21 & 1.07\texttimes{} \\
        URAM with ECC~\cite{xilinx2021basic} & 2.21 & 2.05 & 1.85 & 0.19 & 1.08\texttimes{} \\
        Fixed-point Hamming~\cite{xilinx2021basic} & 2.37 & 2.46 & 2.13 & 0.33 & 0.96\texttimes{} \\
        Unoptimized FFT~\cite{kastner2018parallel} & 2.78 & 2.91 & 2.28 & 0.63 & 0.95\texttimes{} \\
        Multi-stage FFT~\cite{kastner2018parallel} & 2.67 & 2.93 & 2.29 & 0.64 & 0.91\texttimes{} \\
        Huffman encoding~\cite{kastner2018parallel} & 2.63 & 2.32 & 1.99 & 0.33 & 1.13\texttimes{} \\
        Matrix multiplication~\cite{kastner2018parallel} & 2.61 & 2.59 & 2.38 & 0.21 & 1.01\texttimes{} \\
        Parallelized merge sort~\cite{kastner2018parallel} & 2.27 & 2.15 & 1.96 & 0.18 & 1.06\texttimes{} \\
        Vector add with stream~\cite{xilinx2022vitis} & 4.48 & 3.56 & 2.17 & 1.39 & 1.26\texttimes{} \\
        FlowGNN GIN~\cite{flowgnn} & 28.9 & 11.97 & 8.88 & 3.08 & 2.42\texttimes{} \\
        FlowGNN GCN~\cite{flowgnn} & 30.9 & 17.18 & 11.38 & 5.80 & 1.80\texttimes{} \\
        FlowGNN GAT~\cite{flowgnn} & 41.6 & 24.60 & 20.53 & 4.07 & 1.69\texttimes{} \\
        FlowGNN PNA~\cite{flowgnn} & 30.5 & 29.00 & 23.30 & 5.70 & 1.05\texttimes{} \\
        FlowGNN DGN~\cite{flowgnn} & 26.9 & 11.71 & 6.92 & 4.79 & 2.30\texttimes{} \\
        \rowcolor{yellow}INR-Arch~\cite{inrarch} & 128.4 & 26.39 & 9.88 & 16.51 & 4.87\texttimes{} \\
        \rowcolor{yellow}SkyNet~\cite{skynet} & 3103 & 469.3 & 7.55 & 461.8 & 6.61\texttimes{} \\
        \midrule
        \multicolumn{6}{c}{\footnotesize\begin{tabular}{c}
        All times are indicated in seconds.\\
        \textbf{LSv2:} LightningSimV2. \textbf{\thiswork{}} is further broken down into:\\
        \textbf{FE:} front-end compilation (\S\ref{sec:frontend-compilation}). \textbf{MT:} multi-threaded execution (\S\ref{sec:multi-thread-exec}).\\
        $\mathbf{\Delta\downarrow}$\textbf{:} Speedup of \thiswork{} over LightningSimV2.\\
        \end{tabular}}\\
        \bottomrule
    \end{tabular}
    \caption{Comparisons of \thiswork{} over LightningSimV2.}
     \vspace{-20pt}
    \label{tab:compare-lsv2}
\end{table}

Improvements over co-sim are particularly valuable for designs previously unsupported by state-of-the-art simulators, while efficiency remains important. Using LightningSimV2’s own benchmark suite (Table~\ref{tab:compare-lsv2}), we show that \thiswork{} achieves a \textbf{1.26\texttimes{}} geomean speedup over LightningSimV2 with no loss of accuracy. Minor slowdowns observed in four designs (the largest from 2.67 to 2.93 seconds) are likely due to random variations in the runtime environment.

\label{sec:largest-benchmarks}
We draw particular attention to the largest benchmarks listed last in the table, INR-Arch and SkyNet. \thiswork{} reduces the simulation time for INR-Arch by \textbf{4.87\texttimes}, from over 2 minutes to 26.4 seconds, and it reduces SkyNet's simulation time by \textbf{6.61\texttimes} from over 51 minutes to under 8 minutes. Since these Type A designs contain no non-blocking calls and thus must never wait for simulated timing feedback from the shared Perf Sim thread, they observe pure performance gains from our multithreaded architecture. These results show that our flexibly coupled simulation is \textbf{\textit{not a compromise}} over LightningSim's decoupled architecture but actually \textbf{\textit{improves simulation performance}}, especially on the large designs.

\subsubsection{Incremental Simulation}

\begin{table}
    \centering
    \small
    \setlength{\tabcolsep}{2pt}
    \begin{tabular}{l|c|cc|cc|cc}
        \toprule
        \textbf{Description} & \textbf{Depths} & \textbf{Incr.} & \textbf{OK?} & \textbf{FE} & \textbf{MT}& \textbf{Total} & $\mathbf{\Delta\downarrow}$ \\
        \midrule
        Initial run & (2, 2) & --- & --- & 1.75\,s & 0.35\,s & 2.10\,s \\
        Incremental & (2, 100) & 77.86\,\textmu{}s & \textcolor{Green}{\ding{51}} & --- & --- & 77.86\,\textmu{}s & (2.7\textsc{e}4\texttimes) \\
        Non-incremental & (100, 2) & 85.18\,\textmu{}s & \textcolor{red}{\ding{55}} & --- & 0.31\,s & 0.31\,s & (6.77\texttimes) \\
        \midrule
        \multicolumn{8}{c}{\footnotesize\begin{tabular}{c}
        \textbf{Incr.:} Time for incremental re-simulation. \textbf{OK?} if it satisfied constraints (\S\ref{sec:constraints}).\\
        \textbf{FE:} front-end compilation (\S\ref{sec:frontend-compilation}). \textbf{MT:} multi-threaded execution (\S\ref{sec:multi-thread-exec}).\\
        $\mathbf{\Delta\downarrow}$\textbf{:} Speedup over full simulation.\\
        \end{tabular}}\\
        \bottomrule
    \end{tabular}
    \caption{Evaluating \texttt{fig\ref{fig:dataflow-types}\_ex5} under different FIFO depths.}
     \vspace{-16pt}
    \label{tab:incremental}
\end{table}

Like LightningSim, \thiswork{} features the capability to \textit{incrementally} re-simulate designs with changed FIFO buffer depths. To evaluate this capability, we perform a case study on \texttt{fig\ref{fig:dataflow-types}\_ex5}, a Type C design which routes requests between two PEs based on FIFO backpressure.

Results are shown in Table~\ref{tab:incremental}. After the initial run, we re-simulate with each of the two FIFO depths set to 100. In the first case, incremental re-simulation completes in just 78\,\textmu{}s, achieving a \textbf{26,966\texttimes{}} speedup over full simulation. In the second case, incremental simulation fails due to violated \textit{constraints} (Sec.~\ref{sec:constraints}), as the HLS code would have behaved differently under this configuration, requiring a full re-simulation. Even so, reusing the already-compiled executable yields a 6.77\texttimes{} speedup.

\section{Conclusion}

In this work, we proposed \thiswork{}, a novel solution that addresses the simulation limitations of existing HLS tools by bridging the gap between C-level abstractions and hardware-level semantics. \thiswork{} achieves near-C simulation speed while maintaining near-RTL accuracy for both functionality and performance simulation, even for complex dataflow designs.
We first introduced a new taxonomy and conducted an in-depth analysis of dataflow patterns that are particularly challenging to simulate. To tackle these challenges, \thiswork{} carefully orchestrates functionality and performance simulation threads to accurately model hardware-level behavior under arbitrary OS scheduling.
We then demonstrated that \thiswork{} can successfully simulate {\numbenchmarks} designs that were unsupported by existing HLS tools, achieving an average speedup of 32.1\texttimes{} over RTL simulation. We further showed that \thiswork{} provides substantial simulation speedup—up to 6.61\texttimes{}—compared to the state-of-the-art simulator for large-scale designs.
Overall, \thiswork{} enables more agile, productive, and realistic hardware development workflows, bringing HLS significantly closer to being a true RTL replacement.

\bibliographystyle{ACM-Reference-Format}
\bibliography{bibliography}

\end{document}